\documentclass[12pt]{iopart}
\usepackage{epsfig}
\begin{document}

\title[Lattice QCD]{Results from Lattice QCD
\vskip-1.4cm\hfill\small hep-ph/0505073, TIFR/TH/05-14\vskip 1.1cm
}

\author{Rajiv V. Gavai\footnote[3]{E-mail: gavai@tifr.res.in}}

\address{Department of Theoretical Physics, 
Tata Institute of Fundamental Research, \\
Homi Bhabha Road, Mumbai 400 005, India}

\begin{abstract}
I present our recent results on the critical end point in the $\mu_B$-$T$ phase
diagram of QCD with two flavours of light dynamical quarks and compare them
with similar results from other groups.  Implications for a possible energy
scan at the RHIC are discussed.  I also comment briefly on the new results of
great relevance to heavy ion collisions from finite temperature lattice QCD
simulations on speed of sound, specific heat and on the fate of $J/\psi$.
\end{abstract}


\submitto{\JPG}

\maketitle

\section{Introduction}

Lattice Quantum Chromo Dynamics (QCD), defined on a space-time lattice, has
been our best and most reliable tool to extract non-perturbative physics of the
strongly interacting theory for over a decade now. In many cases, such as the
hadron spectra or the weak decay constants, the focus of lattice QCD has now
shifted to obtaining more precise results.  In the fields of strong interest to
this conference, namely Relativistic Heavy Ion Collisions and the possible
transition to the new Quark-Gluon Plasma (QGP) phase, also there are similar
examples : the quark-hadron transition temperature \cite{lp}, $T_c$, the
Wr\'oblewski Parameter \cite{gg}, $\lambda_s$ and the the equation of state
\cite{lp}.  Already a lot of reliable theoretical information on these physical
observables of interest has been predicted by lattice QCD and again the focus
is now for better precision.  Due to the strong influence of dynamical fermions
on thermodynamics and the nature of the phase transition, one still has some
way to go before precise prediction for realistic world of two light ($u$ and
$d$) and one heavy ($s$) flavour become available.  Nevertheless, the
continuing progress in lattice QCD makes it perhaps a reachable goal.  I shall
be unable to review all the important recent lattice results in the time
available. I have therefore chosen to focus on those with thrust on something
new, either conceptually or quantitatively.  In particular, I shall cover the
new results on the $T$-$\mu_B$ phase diagram and the physically interesting
question of $J/\psi$-dissolution/persistence. An interesting theoretical issue
these days has been the comparison of predictions of conformally invariant
theories with lattice QCD results. I shall show one such comparison along with
new results for speed of sound.


\section{QCD Phase Diagram }


The first principles exploration of the QCD phase diagram has the
lattice QCD partition function as its focal point.  Assuming three flavours of
quarks,  $u$, $d$, and $s$ and denoting by $\mu_f$ the corresponding chemical
potentials, it is defined by,
\begin{equation}
{\cal Z} =  \int D U~~ \exp(-S_G)  \prod_{f=u,d,s} 
{\rm Det}~M(m_f, \mu_f)~~.
\label{zqcd}
\end{equation}
Note that the quark mass $m_f$ and the corresponding chemical potential $\mu_f$
enter only through the Dirac matrix $M$ for each flavour.  The baryon and
isospin chemical potentials are related to those for quarks by $\mu_B = \mu_u +
\mu_d + \mu_s$ and $\mu_3 = \mu_u - \mu_d$. 

Progress in unravelling the phase diagram using lattice QCD has been slow due
to so-called ``fermion problem'': for nonzero $\mu$, the determinant in eq.
(\ref{zqcd}) is complex.  This renders both analytical and numerical
investigations very difficult.  Recently, progress was made by a key
observation. Following the phase transition from the $\mu_B = 0$ axis into the
$\mu_B$-$T$ plane is relatively easier.  Various methods have been employed to
achieve this.  In addition to the original re-weighting method  \cite{fk} in
two couplings, imaginary chemical potential \cite{dFP}  and Taylor series
expansion \cite{us} have been used to find and locate the critical end point,
expected for a world with two light flavours on the basis of a variety of model
considerations \cite{models}.  I shall describe below our results obtained by
using the Taylor series method.  One of its primary advantages is the ease of
taking the continuum and thermodynamic limit which is mandatory for all such
lattice computations.  This is especially difficult for the re-weighting method
\cite{fk} due to the $\exp (-\Delta S)$ factor it has which becomes
exponentially small for bigger lattices.  Furthermore, the discretization
errors propagate in an unknown manner in a re-weighting computation, making
reliable estimates in continuum limit very difficult. A much better control of
the systematic errors is, on the other hand, achieved in the method we employ.
We study volume dependence of the Taylor coefficients at several temperature
$T$ to i) bracket the critical region and then to ii) track its change as a
function of volume. As shown in Figure 1, one expects to pin down the critical
region much more precisely this way in going from a smaller volume $V_1$ to a
bigger one $V_2$.

\begin{figure}[htb]
\begin{center}
\epsfxsize=6.4cm
\epsfbox{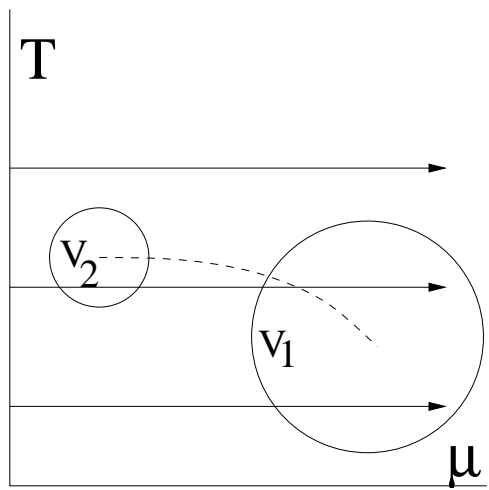}
\label{fig1}
\caption{ Volume dependence of the critical region.}
\end{center}
\end{figure}
\noindent 
Various quark, baryon and isospin densities and the
corresponding susceptibilities can be obtained from eq. (\ref{zqcd}) as:  

\begin{equation}
\qquad \qquad n_i =  \frac{T}{V} {{\partial \ln {\cal Z}}\over{\partial \mu_i}}, \qquad 
\chi_{ij} =  \frac{T}{V} {{\partial^2 \ln {\cal Z}}\over{\partial \mu_i 
\partial \mu_j} }  ~~.
\label{nchi}
\end{equation}
Higher order susceptibilities,  $\chi_{n_u,n_d}$, can be similarly defined as
$n_u$($n_d$) partial derivatives of $ \ln {\cal Z}$ with respect to
$\mu_u$($\mu_d$). Using them, the pressure $P$ has the following expansion in
$\mu$:
\begin{equation}
   \frac{\Delta P}{T^4} \equiv \frac{P(\mu, T)}{T^4} - \frac{P(0, T)}{T^4}
   = \sum_{n_u,n_d} \chi_{n_u,n_d}\;
        \frac{1}{n_u!}\, \left( \frac{\mu_u}{T} \right)^{n_u}\,
        \frac{1}{n_d!}\, \left( \frac{\mu_d}{T} \right)^{n_d}\,
\label{nnchi}
\end{equation}

Quark number susceptibilities in \eref{nchi} are crucial for many quark-gluon
plasma signatures which are based on fluctuations in globally conserved
quantities such as baryon number or electric charge.  Theoretically, they serve
as an important independent check on the methods and/or models which aim to
explain the large deviations of the lattice results for pressure $P$($\mu$=0)
from the corresponding perturbative expansion.  This has been discussed
elsewhere in details. Here we will be concerned with using them with the higher
order terms to learn about the critical end point.  Indeed, a series for
baryonic susceptibility $\chi_B$ can be constructed using them, whose radius of
convergence gives the nearest critical point \cite{us1}.  Limited by the finite
(and even small) number of terms available, one uses standard tests for series
convergence on them to do the best one can. Successive estimates for the radius
of convergence can be obtained from these using $r_{n+2} =
\sqrt{{\chi^n_B}/{\chi^{n+2}_B}}$.  We use terms up to 8th order in $\mu$,
i.e., the estimates from 2/4, 4/6 and 6/8 terms in eq.(\ref{nnchi}).  It may be
noted here that a similar construction of a series goes through for the
off-diagonal susceptibility, $\chi_{11}$ as well. The ratio
$\chi_{11}/\chi_{20}$ can be shown to yield the ratio of widths of the measure
in the imaginary and real directions at $\mu=0$, i.e., a measure of the 
difficulty of the `fermion sign' problem mentioned above.  With some care, 
this argument can be generalized to nonzero $\mu$. 

\subsection{Our Results}

\begin{figure}[htb]
\begin{minipage}{0.49\textwidth}
\epsfxsize=6.5cm
\epsfbox{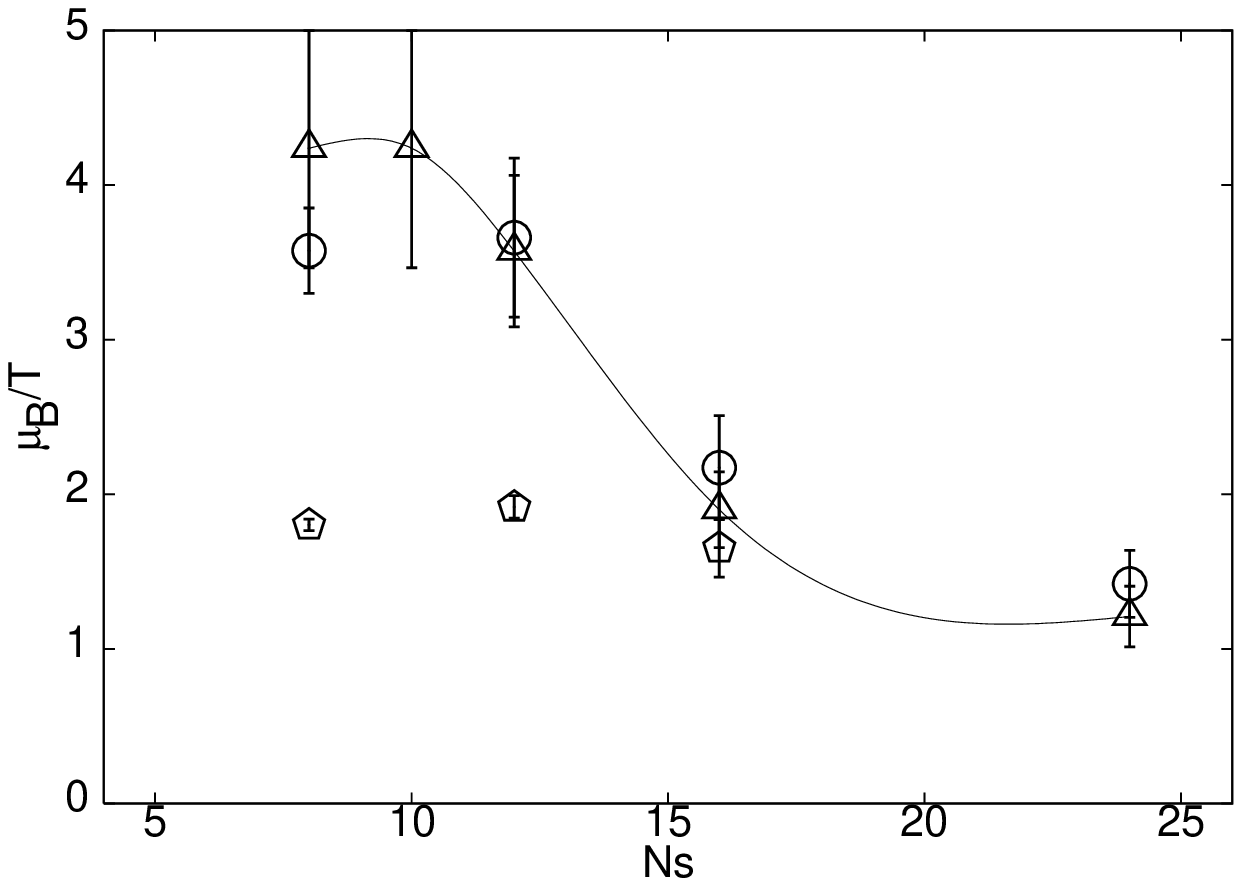}
\end{minipage}
\begin{minipage}{0.49\textwidth}
\epsfxsize=6.4cm
\epsfbox{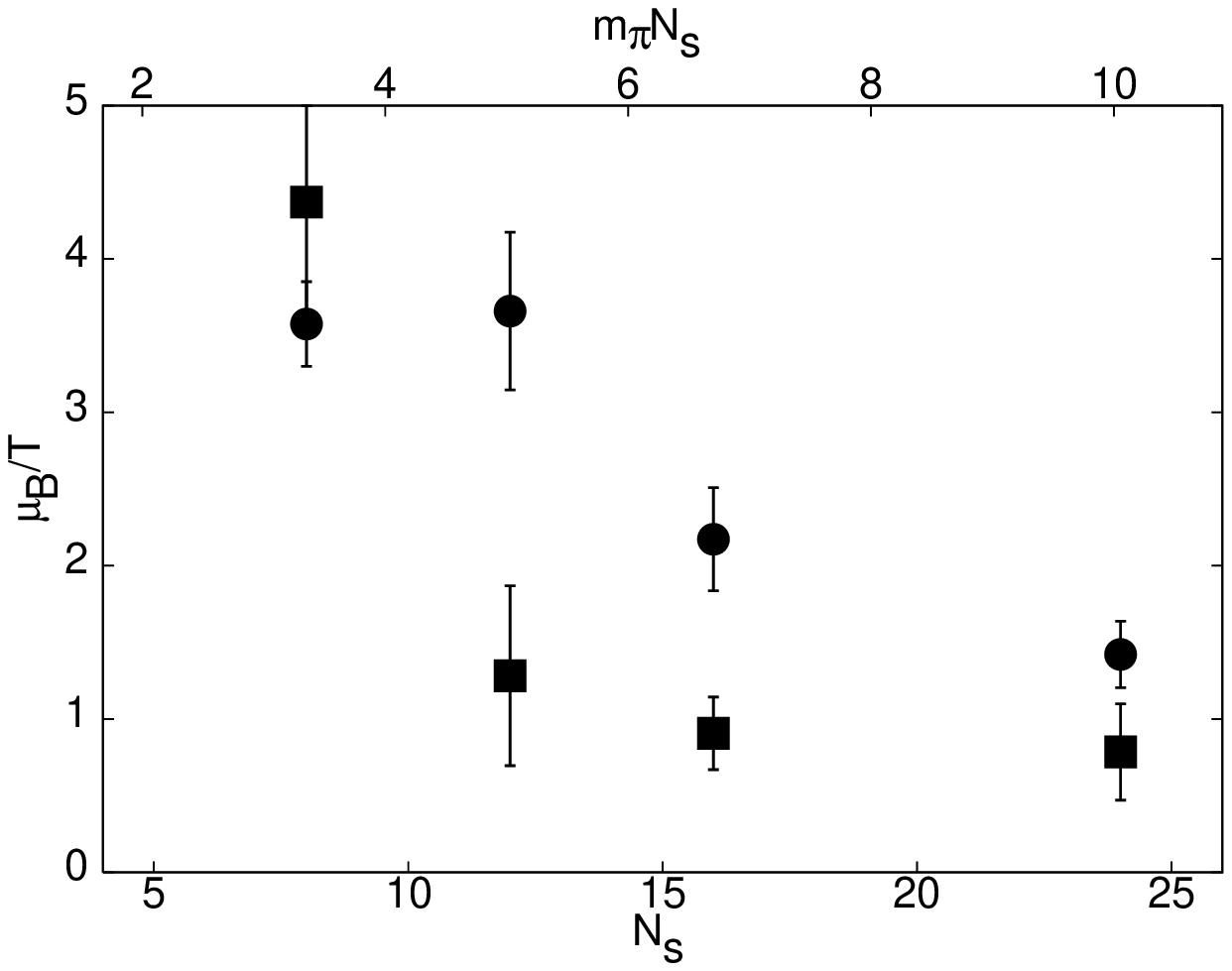}
\end{minipage}
\label{fig2}
\begin{center}
\caption{The radius of convergence as a function of $N_s$ ($m_\pi N_s$). 
   The left panel shows results obtained by comparing the 4th and 6th order
   at $T/T_c=0.9$ (pentagons), 0.95 (circles) and 1 (triangles). The
   right panel shows results obtained at $T/T_c=0.95$ using the ratio of
   the 4th and 6th order NLS (boxes) and the 6th and 8th order NLS (circles).}
\end{center}
\end{figure}

Our results \cite{us1} were obtained by simulating full QCD with two flavours
of dynamical staggered quarks of mass $m/T_c =0.1$, where $T_c$ is the
transition temperature for zero chemical potential, on lattices of size $4
\times N_s^3$, with $N_s$ = 8, 10, 12, 16 and 24. R-algorithm \cite{gott} with
a trajectory length of one unit of molecular dynamics time on $N_s$ = 8 was
used for simulations of full QCD.  The trajectory length was scaled with $N_s$
on larger spatial lattices.  Earlier investigations have pinned down the
physical parameters for our simulations at zero chemical potential \cite{milc}
: $m_\rho/T_c = 5.4 \pm 0.2$ and $ m_\pi/m_\rho =0.31 \pm 0.01$.  We used the
canonical methods to fix the lattice spacing $a$ in physical units for which we
made runs on symmetric $16^4$ lattices.  Their details  as well as other details
about evaluation of the higher order susceptibilities can be found in
\cite{us1}.  Here I shall be brief and state only that our simulations were
made at $T/T_c$ = 0.75(2), 0.80(2), 0.85(1), 0.90(1), 0.95(1), 0.975(10),
1.00(1), 1.05(1), 1.25(1), 1.65(6) and 2.15 (10) with a typical statistics of
50-100 in the units of (maximum) autocorrelation length.

Using terms of up to 8th order in $\mu$ in the expansion in eq. (\ref{nnchi}),
we obtained estimates for the radius of convergence shown in Figure 2.  The
left panel shows $r_6$ as a function of $N_s$ at three temperatures whereas the
right panel exhibits $r_6$ and $r_8$ at $T/T_c =$ 0.95. The latter also shows
on top the spatial lattice size in units of inverse pion length.  As seen from
the left panel of the figure, our estimate  of the radius of convergence for
the smaller lattices with $N_s \sim 10$ is consistent with the critical end
point estimate of the re-weighting method \cite{fk} with appropriate size, i.e,
$N_s m_\pi$ = 3-4. Note that  $m_\pi/m_\rho = 0.31$ in that case as well.
Figure 2 shows, however, large finite size effects for these small lattices and
a strong change in the estimate around $N_s m_\pi \sim 6$.  Such finite size
effects are expected theoretically \cite{lm} and have been seen in other cases
such as chiral condensates\cite{gr}  and lattice QCD evaluations of structure
functions \cite{jan}.   A key consequence of our use of larger volumes than
before is that we find a shift in the critical end point to $\mu_B/T \sim 1-2$,
as is evident from Figure 2. Recently, similar computations but up to 6th order
have been reported on a large lattice with $N_s m_\pi \sim 15$, and were also
discussed in this conference \cite{bs}.  These use, however, large quark
masses, and consequently a a large $m_\pi/m_\rho \sim 0.7$.   It will be
interesting to see if lowering their quark mass leads to results similar to
ours.   

\subsection{Measure of Sign Problem}

\begin{figure}[htb]
\begin{center}
\epsfxsize=6.4cm
\epsfbox{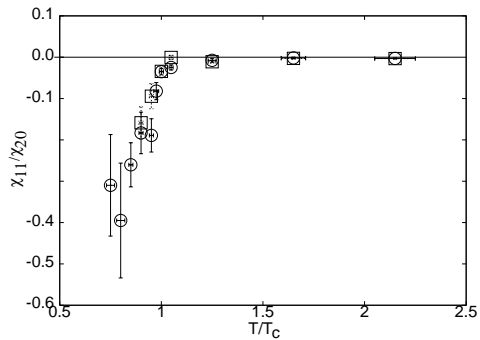}
\label{fig3}
\caption{The ratio of quark number susceptibilities $\chi_2$ and $\chi_{11}$
on $4\times16^3$ (circles) and $4\times24^3$ (boxes) lattices.  }
\end{center}
\end{figure}

Figure 3 displays the ratio of the two susceptibilities $\chi_{11}$ and
$\chi_{20}$ as a function of temperature on two different lattice sizes.  This
ratio, which is a measure of the seriousness of sign problem, shows how the
increase in both its value and fluctuations make the simulations progressively
more difficult as one lowers the temperature to $T_c$ and below.  The success
of the current round of finite density lattice simulations can also be traced
from this figure to the fact that all current efforts, regardless of the
details of their methods, attempt to explore the phase diagram starting from
$T_c$, the transition point at zero chemical potential.  An important check one
can make in the Taylor series expansion technique is that each coefficient in
the Taylor expansion must be volume independent.  Since the actual evaluation
of these terms entails cancellation of divergences, this is indeed a nontrivial
check which has been made successfully \cite{us1}.  In fact, care is needed in
extracting any equation of state at finite chemical potential from these
coefficients : a systematic study on different volumes (spatial lattices) with
a subsequent extrapolation to infinite volume is necessary to ensure that some
peak-like behaviour in, e.g., $\chi_{40}$, is real and does not give spurious
results for thermodynamic quantities.

\section{Speed of Sound}

\begin{figure}[htb]
\begin{minipage}{0.49\textwidth}
\epsfxsize=6.5cm
\epsfbox{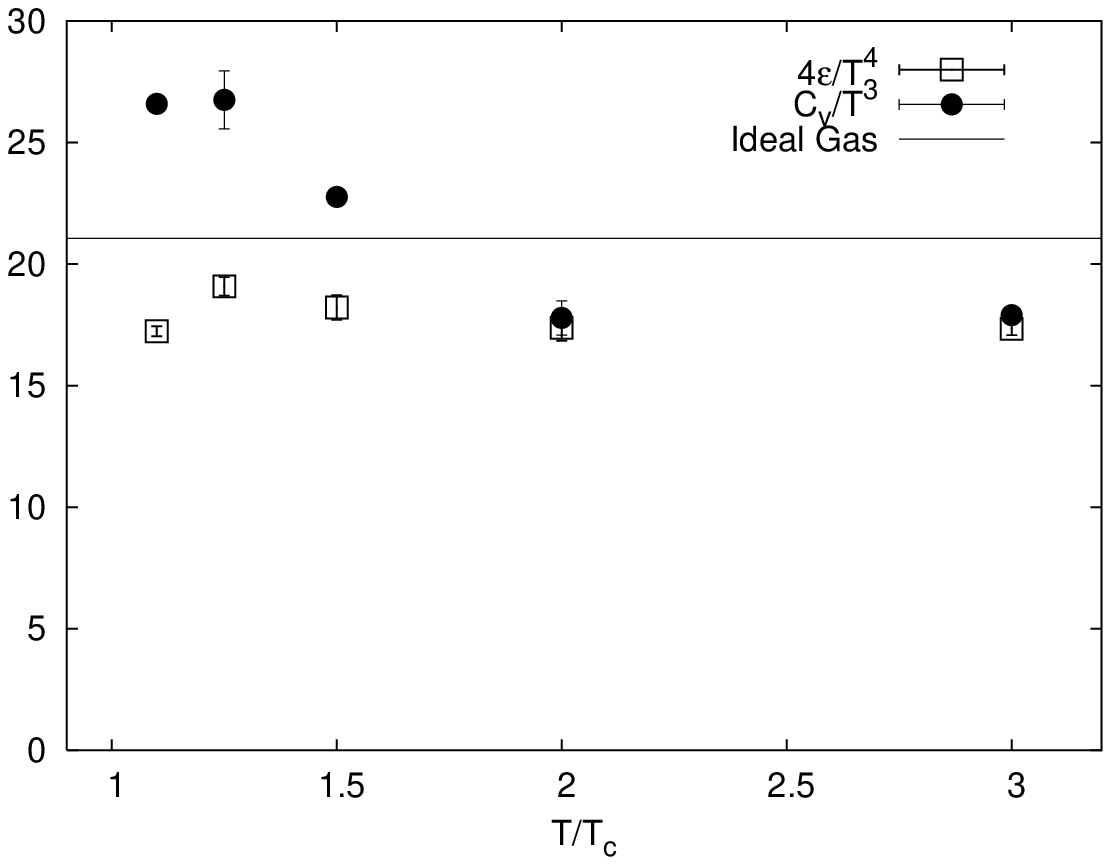}
\end{minipage}
\begin{minipage}{0.49\textwidth}
\epsfxsize=6.4cm
\epsfbox{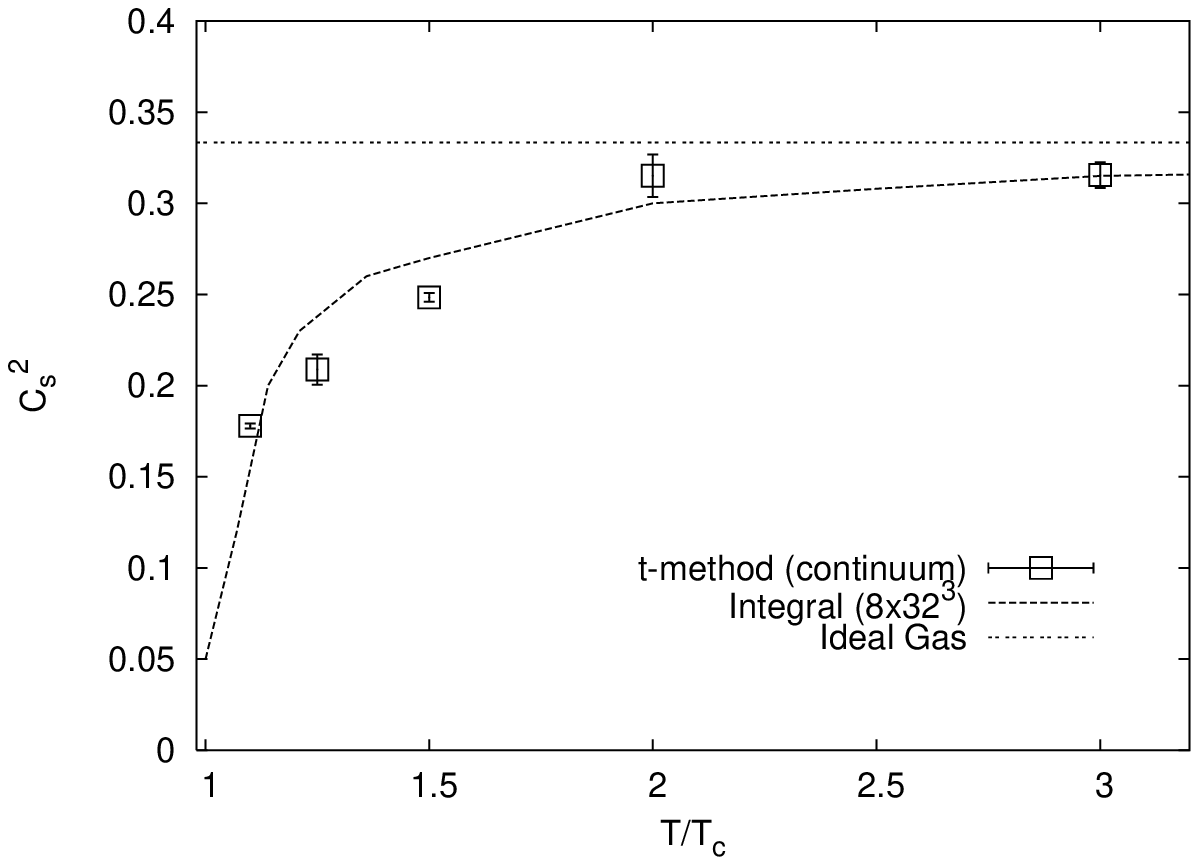}
\end{minipage}
\label{fig4}
\begin{center}
\caption{Specific heat $C_v$ and 4$\epsilon$ (left) and the speed of sound
$c_s$ as a function of temperature for quenched QCD. See text for more details.}
\end{center}
\end{figure}

An important physical property of the strongly interacting matter at finite
temperature, which has been computed \cite{old} since long using lattice
techniques, is the the speed of sound.  Its relevance to many aspects of heavy
ion physics, such as the elliptic flow or the hydrodynamical studies can hardly
be overemphasized.  A related quantity of interest is the specific heat which
has been related to event-by-event fluctuations of the transverse momentum 
($p_T$).  There are various measures of these fluctuations currently in use 
in the literature.  It may actually be better to relate them to the 
specific heat in order to make them both more physical as well as 
uniform in convention.

A recent development in this direction on the lattice is a new way to obtain
these by relating them to the temperature derivative of an anomaly measure
$\Delta/\epsilon \equiv (\epsilon -3 P)/\epsilon$ \cite{swag1}.  Using lattices
with 8, 10, and 12 temporal sites ($38^3 \times 12$ and $38^4$ lattices) and
with statistics of 0.5-1 million iterations, $\epsilon$, $P$, $s$, $C_s^2$ and
$C_v$ were obtained in continuum using this differential method at 2$T_c$ and
3$T_c$ for quenched QCD.  Sizable deviations from the corresponding ideal gas
results were found although $C_v$ was consistent with 4$\epsilon$, as expected
in a weak coupling scenario.  On the other hand the result for entropy density
at 3$T_c$ also seemed to agree with a strong coupling (super Yang-Mills)
prediction \cite{sym} : $s/s_0 = f(g^2 N_c)$, where
$f(x)=\frac34+\frac{45}{32}\zeta(3)(2x^{-3/2})+\cdots$ and $s_0 = \frac23 \pi^2
N_c^2T^3$.  However, curiously it too failed at 2$T_c$.

New results for the specific heat and the speed of sound, obtained by using a
new method, called the $t$-method, were presented at this conference in a
poster \cite{swag2}.  While the differential method \cite{bidif} for obtaining
thermodynamic quantities has lesser systematic errors and is convenient to use,
it was abandoned in favour of the integral method \cite{int} due to the
former's problem of yielding negative Pressure.  It was shown that this problem
of negative pressure can be completely avoided even in the differential method
by writing the derivatives with respect to temperature and volume using a
parameter $t$ : the old (Bielefeld) method \cite{bidif} corresponds to $t = 1$
and leads to negative pressure but $t =0 $ results in positive pressure always.
Figure 4 shows these new results of \cite{swag2} for the specific heat (the
left panel) and the speed of sound (right panel).  While the corresponding
pressure is not shown here, it is positive even around $T_c$, an input which
went in getting the speed of sound (squares), shown in Figure 4. Note that
these continuum results have been obtained by simulating lattices with many
lattice spacings and then extrapolating to zero unlike the results for the
integral method (dotted line) in Figure 4 which are for $N_t =8$.  The
interesting and noteworthy features are : i) the specific heat does tend to
peak near $T_c$ and is widely different there from the weak coupling
expectation of 4$\epsilon$, ii) for $ T \ge 2 T_c$, this weak coupling
expectation is fulfilled but the resulting value is very different from that
for an ideal gas, iii) speed of sound is about 10 \% smaller than the ideal gas
for higher temperatures but drops close to $T_c$, and iv) the results are in
good agreement with the integral method but could differ significantly in the
neighbourhood of $T_c$.  Further investigations are needed to see if $c_s^2$
dips close to zero near $T_c$.

\section{Persistence of $J/\psi$}

\begin{figure}[htb]
\begin{minipage}{0.49\textwidth}
\epsfxsize=6.5cm
\epsfbox{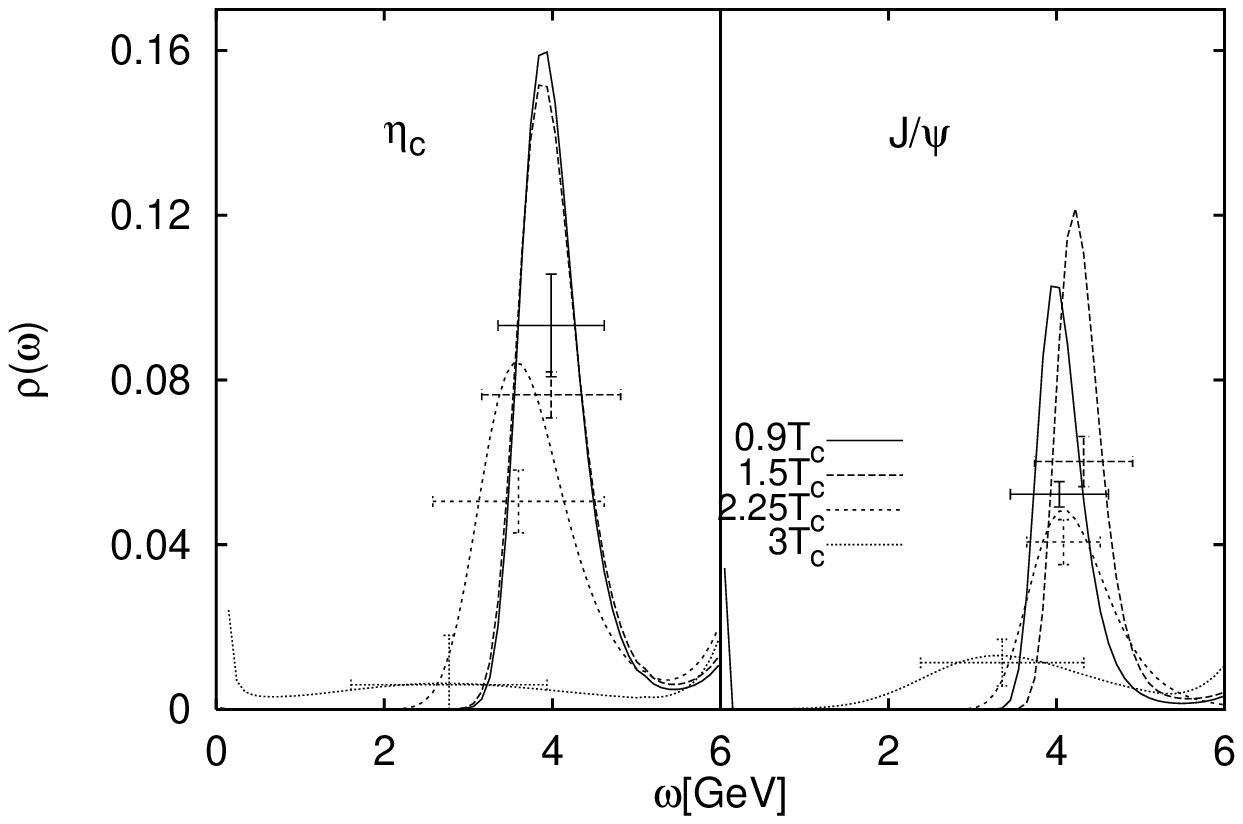}
\end{minipage}
\begin{minipage}{0.49\textwidth}
\epsfxsize=6.4cm
\epsfbox{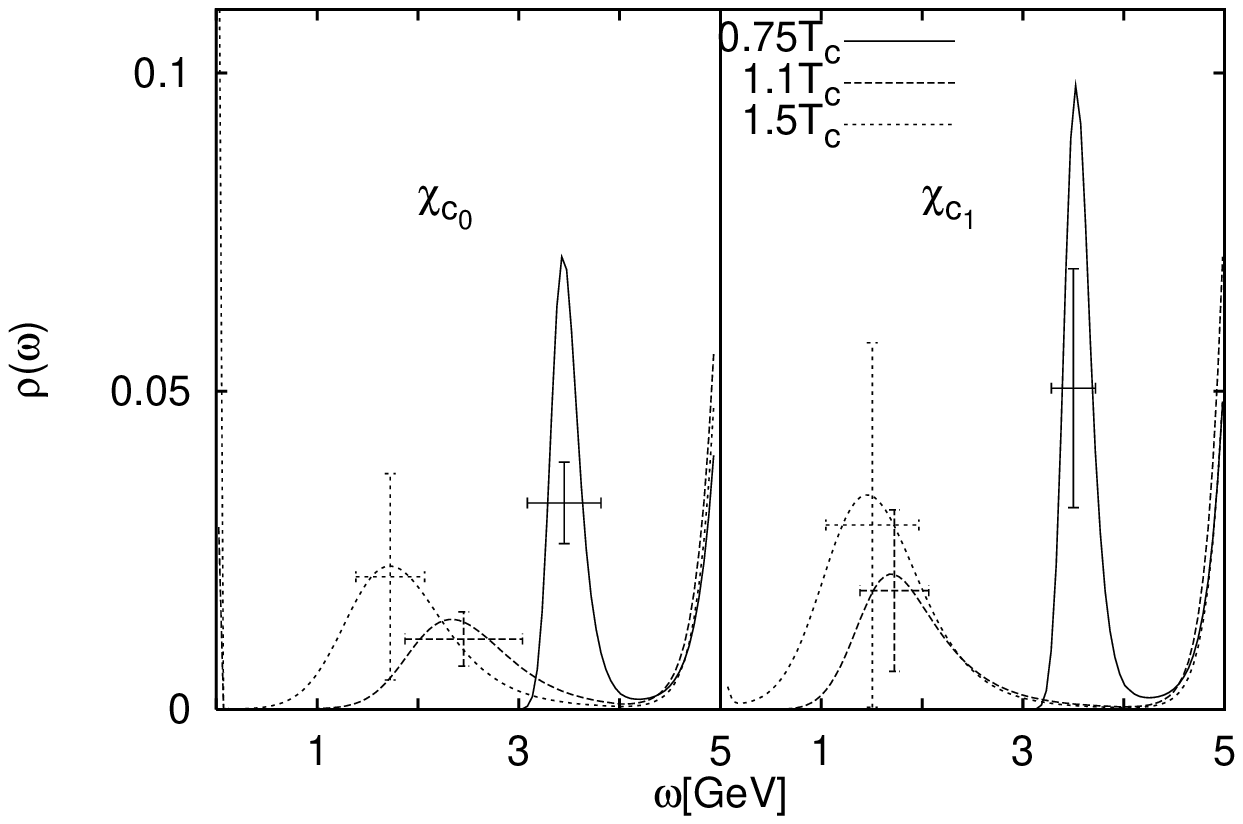}
\end{minipage}
\label{fig5}
\begin{center}
\caption{Spectral functions \cite{saumen} of the vector $J/\psi$ and scalar
$\chi$ mesons as a function of energy at various temperatures. See text for
more details.}
\end{center}
\end{figure}

An intriguing result on the $J/\psi$ and similar heavy quarkonia has recently
emerged \cite{saumen,japs} from the lattice QCD investigations. As is well
known,  Matsui and Satz \cite{ms} proposed $J/\psi$ suppression as a signal of
quark-gluon plasma.  Quarkonium potential model calculations with an ansatz for
temperature dependence of the heavy quark-antiquark potential lead to the
result that $J/\psi$ and $\chi_c$ states dissolve in QGP by 1.1$T_c$.  The
impressive results from the CERN NA50 \cite{na50} did suggest an anomalous
suppression of $J/\psi$ in Pb-Pb collisions, bringing forth detailed
theoretical questions on this suppression mechanism and the above mentioned
ansatz.  In particular, one approach taken by the lattice experts
\cite{saumen,japs} has been to study the spectral function of these states in
QGP to look for their dissolution from first principles.  The recognition of
the maximum entropy method (MEM) as a reliable tool to extract the spectral
functions from temporal lattice correlators at zero temperature lead to its
application at finite temperature.  Caution is, however, called for in doing so
since nonzero and high temperatures are obtained by making the temporal lattice
smaller whereas MEM works better for longer temporal lattices. 

Figure 4 shows the results of the Bielefeld group \cite{saumen}, obtained by
simulating lattices from $48^3 \times 12$ to $64^3 \times 24$ in sizes in the
quenched approximation of no dynamical quarks.  The vertical bars are drawn to
indicate the size of the error on the integrated spectral function in the area
shown by the horizontal bars.  One observes that the $\chi_c$ does seem to
indeed dissolve by 1.1$T_c$, as expected from naive potential model
considerations.  However, the $J/\psi$ and $\eta_c$ persist up to 2.25 $T_c$
and are gone only at 3$T_c$.  Similar results have been obtained by other
groups \cite{japs} as well, although there seems to be some disagreement on the
precise value of the temperature at which the $J/\psi$ peak vanishes.

While these results have to be still confirmed in a full QCD simulation with
light dynamical quarks, they already can have strong consequences for the
$J/\psi$ signal.  Recall that only about 30-40 \% of the $J/\psi$ come through
$\chi$ decays ( with a substantial fraction from $\chi_2$ which has not yet
been investigated ), thus the predicted suppression will be rather small until
temperatures close to 3$T_c$ are reached (or energy densities 81 times that of
critical energy density is reached), which seems to be a tall order for the
current round of CERN and BNL experiments.  An interesting observable
consequence could, however, be a change of the suppression pattern as a
function of $\sqrt{s}$ or the energy density reached.

\section{Summary}

The QCD phase diagram in $T-\mu_B$ plane has begun to emerge. Since all
results so far have been obtained on $N_t = 4$, i.e, very coarse, lattices,
several quantitative issues related to continuum limit still have to wait.
It may be noted that the existence of equally legitimate \cite{mu} different 
prescriptions of introduction of the chemical potential on the lattice
can, and do, lead to even larger lattice artifacts on such coarse lattices.
Nevertheless, it is very heartening to find that different methods lead to a 
similar qualitative picture.

Our results, using Taylor expansion, permit us to take the thermodynamic limit,
unlike other methods.  Our investigation of volume dependence suggests $N_s
m_\pi > 6 $ approximates the thermodynamic volume limit reasonably.  The
$\mu_B/T$ of the critical end point, identified from the radius of convergence
using up to, and including the 8th order terms, drops strongly around that
volume.  We find  that $\mu_B/T \sim 1-2$ is indicated for the critical point
with a corresponding temperature of $\simeq 0.95 T_c$, where $T_c$ is the
transition temperature for the baryonfree case.  

\begin{figure}[htb]
\begin{center}
\epsfxsize=8.5cm
\epsfbox{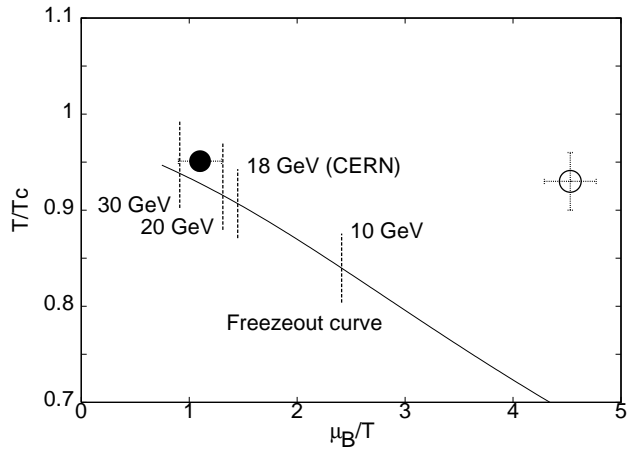}
\label{fig6}
\caption{The QCD phase diagram with a freeze-out curve superimposed. The
   filled circle denotes the estimate of the critical point which has been
   obtained in \cite{us1}. The open circle is an earlier estimate from
   \cite{fk} using smaller lattices and nearly the same quark mass.}
\end{center}
\end{figure}

Figure 6 shows our results, together with the earlier determination \cite{fk}
on small volumes but similar quark mass, i.e, the pion mass.  A freeze-out
curve \cite{jean}, obtained by treating a resonance gas as ideal, has been
superimposed on the Figure.  From the marked values of the CM energy, $\sqrt
s$, per nucleon needed to reach that point on the curve,  one notes that the
critical end-point may be in the reach of future RHIC energy scans.

I also presented new continuum results on the speed of sound and the specific 
heat for quenched QCD.  A new method, which revives the earlier
differential method but without its drawback of negative Pressure was used to
obtain these results. Interesting quantitative agreement was found with
strong coupling predictions from super Yang-Mills theory at 3$T_c$ but not at
2$T_c$.

Finally, the intriguing persistence of $J/\psi$ in QGP was discussed. The MEM
technique has been used to obtain the spectral function at finite temperature.
While it showed the peak in the $\chi$ spectral function to go away by
1.1$T_c$, the peak in the vector spectral function persists up to 2.25-3$T_c$.
Extension of these simulations to incorporate dynamical quarks would be crucial
to be sure of the phenomenological consequences of these results.

\vspace{0.5cm}


\end{document}